\documentclass[12pt]{article}

\pdfoutput=1

\usepackage{a4wide}
\usepackage{amssymb}
\usepackage{graphicx}
\usepackage{verbatim}
\usepackage{textcomp}

\usepackage[intlimits]{amsmath}
\usepackage{amsfonts}

\usepackage{epsfig}

\newcommand{\be}{\begin{equation}}
\newcommand{\ee}{\end{equation}}
\newcommand{\bea}{\begin{eqnarray}}
\newcommand{\eea}{\end{eqnarray}}
\newcommand{\ba}{\begin{eqnarray}}
\newcommand{\ea}{\end{eqnarray}}

\newcommand{\WW}{{\cal W} }

\def\XXint#1#2#3{{\setbox0=\hbox{$#1{#2#3}{\int}$}
     \vcenter{\hbox{$#2#3$}}\kern-.52\wd0}}

\begin{document}

\setcounter{table}{0}

%\begin{flushright}\footnotesize
%\texttt{ICCUB-20-022}
%\end{flushright}

\mbox{}
\vspace{0truecm}
\linespread{1.1}

\vspace{0.5truecm}

\centerline{\Large \bf  Phantoms and strange attractors in cosmology} 

%\medskip
%\centerline{\Large \bf and large $N$ phase transitions} 

\vspace{1.3truecm}

\centerline{
    {\large \bf Jorge G. Russo} }

\vspace{0.8cm}

\noindent  
\centerline {\it Instituci\'o Catalana de Recerca i Estudis Avan\c{c}ats (ICREA), }
\centerline{\it Pg. Lluis Companys, 23, 08010 Barcelona, Spain.}

\medskip
\noindent 
\centerline{\it  Departament de F\' \i sica Cu\' antica i Astrof\'\i sica and Institut de Ci\`encies del Cosmos,}
\centerline{\it Universitat de Barcelona, Mart\'i Franqu\`es, 1, 08028 Barcelona, Spain. }

\medskip

\centerline{  {\it E-Mail:}  {\texttt jorge.russo@icrea.cat} }

\vspace{1.2cm}

\centerline{\bf ABSTRACT}
\medskip

We study a cosmological model of  gravity coupled to three, self-interacting scalar fields, one of them with negative kinetic term.
The theory has cosmological solutions described by three-dimensional quadratic autonomous equations.
%, leading to  strange attractors.
Remarkably, the dynamical system has strange attractors, which are in fact very similar to the classic Lorenz attractor.
 The associated chaotic cosmologies exhibit highly fluctuating periods of contraction and expansion, alternating with long, steady periods  in a de Sitter-like phase.

\noindent

\vskip 1.2cm
%\noindent {Keywords: Matrix model, large N, Penner's model}
\newpage

\def\sech{ {\rm sech}}
\def\p{\partial}
\def\pa{\partial}
\def\ov{\over }
\def\a{\alpha }
\def\g{\gamma}
\def\s{\sigma }
\def\td{\tilde }
\def\vp{\varphi}
\def\strokedint{\int}
\def \ha {{1 \over 2}}

\def\KK{{\cal K}}

%\newcommand\cev[1]{\overleftarrow{#1}} 

%%\renewcommand{\thefootnote}{\fnsymbol{footnote}}
%%\setcounter{footnote}{0}
%%%%%%%%%%%%%%%%%%%%%%%%%%%%%%%%%%%%%%%%%%%%%%%%%%%%%%%%%%%%%%%%%%%%%

%\clearpage

\textwidth = 460pt
\hoffset=0pt

%\tableofcontents

%%%%%%%%%%%%%%%%%%%%%
%\bigskip

%%%%%%%%%%%%%%%%%%%%%%%%%%%%%%
\section{Introduction}

Chaos in cosmology arises naturally and has been widely discussed
in many different contexts and models \cite{Barrow:1981sx,Belinski:2014kba}.
A plausible reason behind the chaos is the influence of arbitrary perturbations  that become uncontrollable near the cosmological singularity, giving rise to complicated, non-homogeneous geometries. 
These can  initially be described by generalized Kasner metrics, but  soon acquire a  strong oscillatory behavior that becomes stochastic \cite{Belinski:2014kba}. The classical description fails as
physics enters the unknown quantum realm of gravity.

Historically, chaos  appeared in a simplified  model of  atmospheric convection proposed by the mathematician and meteorologist Edward Lorenz in 1963.
The model gave rise to a 
 three-dimensional quadratic autonomous system,
 whose detailed analysis led to the discovery of strange attractors. The cosmological model studied here is closely related to the Lorenz model.
 %; indeed, it is a simple generalization, as we shall see.

Here the analysis of chaos will be greatly simplified because we will be working with first-order equations. The idea is to construct cosmological models
for multiscalar fields interacting with gravity, where the scalar potential can be
derived from a (fake) superpotential. This plays the role of a Hamilton's characteristic function of Hamilton-Jacobi theory.
The method is standard (a discussion can be found in \cite{Skenderis:2006rr}).
Starting with a Lagrangian
\be
L =\frac12 G_{ij} \partial_t \phi^i \partial_t \phi^j-V(\phi^i)\ ,
\ee
one assumes that a superpotential $W$ exists, satisfying
\be
\label{potiuno}
V=-\frac12 G^{ij} \frac{\partial W}{\partial \phi^i}\frac{\partial W}{\partial \phi^j}\ .
\ee
Then, the solutions of the first order system
\be
\label{ecuuno}
\partial_t \phi^i =\epsilon G^{ij} \frac{\partial W}{\partial \phi^j}\ ,
\ee
with $\epsilon=\pm 1$,  solve the second-order equations.
In general, there may be solutions of the second-order equations that are not described by the first-order system, but our aim here
is to discuss specific solutions.

Cosmological models described in terms of trajectories in a three dimensional space
have first appeared in \cite{Russo:2022pgo}, in the context of the ``universal" one-axion model
(generalizing the model of \cite{Sonner:2006yn}
where cosmologies were associated to trajectories in a 2d space). In the case of \cite{Russo:2022pgo}, too, the dynamics is governed by first-order equations, although this reduction to a first-order system occurs in a completely different way (for other approaches, see {\it e.g.}  \cite{Odintsov:2017tbc}). The three-dimensional autonomous system of \cite{Russo:2022pgo} is never chaotic in any range of the parameters. An important property of the axion-two dilaton system of \cite{Russo:2022pgo} is that, for specific values of the parameters, represents
a consistent truncation of massive maximal supergravities.
While the scalar field potential for the present models is motivated by closely related potentials 
appearing in 
D-brane lagrangians, it differs from these in an essential way.
%,as discussed below. 

\medskip

\section{The non-chaotic D-brane model}

The scalar field potential that we will choose for our cosmological model is
motivated by a closely related potential that appears in
some supersymmetric Yang-Mills theories.
Cosmological models based on similar potentials  with matrix-valued scalar fields have been investigated in the past \cite{Ashoorioon:2009wa}.
More recently, there has also been  interest in chaos in matrix models and Yang-Mills theories, motivated by the holographic description of black hole
horizons and the onset of thermalization (see {\it e.g.} \cite{Asano:2015eha,Gur-Ari:2015rcq,Baskan:2019qsb,Baskan:2022dys}). 

Here we shall consider the ${\cal N}=1^*$ theory that arises by adding
mass terms to the three chiral multiplets $\hat \Phi_{i}$, $i=1,2,3$, of the ${\cal N}=4$ $SU(N)$ super Yang-Mills theory. It has a superpotential
\be
W_{{\cal N}=1} = 
%\frac{2\sqrt{2}}{g_{\rm YM}^2} 
{\rm Tr}\left( [\hat \Phi_1,[\hat \Phi_2,\hat\Phi_2]] +m_1\hat \Phi_1^2 +m_2\hat\Phi_2^2+m_3\hat\Phi_3^2 \right)\ .
\ee
The $\hat \Phi_{123}$ are in the adjoint representation of $SU(N)$. Let us denote by $\varphi_{i}$,  $i=1,...,6$ the six (real) scalar components. The bosonic part of the Lagrangian containing the scalar fields is
\be
L={\rm Tr}\left(-\frac12
D_\mu \varphi_i D^\mu \varphi_i - \frac12 m_i^2 \varphi_i \varphi_i +\frac14 [\varphi_i,\varphi_j][\varphi_i,\varphi_j]\right)\ .
\ee
We shall consider the ansatz
\bea
&& Z_1 \equiv \varphi_1+i\varphi_4= x(\sigma) e^{-i\mu_1 t}\alpha_1\ ,
\nonumber\\ 
&& Z_2 \equiv\varphi_2+i\varphi_5= y(\sigma) e^{-i\mu_2 t}\alpha_2\ ,
\nonumber\\ 
&& Z_3 \equiv \varphi_3+i\varphi_6= z(\sigma) e^{-i\mu_3 t}\alpha_3\ ,
\eea
where $\sigma $ is one of the spatial coordinates and $\alpha_{123}$ are any $N\times N$  representation of the $SU(2)$ algebra,
\be
\big[\alpha_i,\alpha_j \big]=2i\epsilon_{ijk}\alpha_k \ ,\qquad {\rm Tr}[\alpha_i\cdot\alpha_j] =a_N\, \delta_{ij}\ .
\ee
For irreducible representations, $a_N=\frac13 N(N^2-1)$; for reducible representations, the coefficient analog to $a_N$ is smaller.
We assume a configuration where the gauge field vanishes. Note that the Gauss law constraint 
\be
\sum_i\left[\partial_\mu \varphi_i,\varphi_i\right]=0\ ,
\ee
is identically satisfied by the given ansatz.

The resulting equations for $x$, $y$, $z$ can be derived from the effective Lagrangian
\be
L_{\rm eff} = 
 -\frac12 
 %\, a_N
 \Big(( \partial_\sigma x)^2+ ( \partial_\sigma y)^2 + ( \partial_\sigma z)^2+ M_1^2 x^2+ M_2^2 y^2+M_3^2 z^2 \Big)
%\nonumber \\
%&+& 
-2 \big(
x^2 y^2 +x^2 z^2 + y^2 z^2\big)
 ,
\ee
where 
\be
\quad M_i=\pm \sqrt{-\mu_i^2+m_i^2}\ .
\ee
The  second-order equations for $x(\sigma),\ y(\sigma),\ z(\sigma)$
are, in general, difficult to solve,
but, when $M_1+M_2+M_3=0$,  one can write first-order equations using the following superpotential
\be
W=\frac12 (M_1 x^2+M_2 y^2+M_3 z^2)+2 xy z\ .
\ee
It satisfies
\be
V=\frac12 (\partial_x W\partial_x W+\partial_y W\partial_y W+\partial_z W\partial_z W)\ ,
\ee
where
$$
V\equiv \frac{1}{2}(M_1^2 x^2+M_2^2 y^2+M_3^2 z^2)+2(x^2y^2+x^2z^2+y^2z^2)\ ,
$$
provided
\be
M_1+M_2+M_3=0\ .
\ee

The associated first-order equations (\ref{ecuuno}) are, therefore,
\be
\label{didi}
 x' = M_1 x+2 y z\ ,\quad y' = M_2 y +2 xz\ ,\quad  z' = M_3 z +2 xy\ . 
\ee
%This three-dimensional dynamical system is not chaotic.

\section{Chaos with ghosts}

While the dynamical system (\ref{didi}) is not chaotic, it is nevertheless closely related to a chaotic system discovered in \cite{Liuchen,Lucheng}, by looking
for generalizations of the classic Lorenz chaotic system.
The equations are 
%(see also \cite{Luchen})
\be
\label{lulu}
 x' = a x+ y z\ ,\quad y' = b y + xz\ ,\quad  z' = c z - xy\ . 
\ee
%with $c=-ab/(a+b)$.
The difference  with (\ref{didi}) lies in the opposite sign in the quadratic term of the $z$ equation, and it turns out to be a crucial difference (the extra factor of 2 in the quadratic terms in  (\ref{didi}) is not important as it can be rescaled away). As discussed below, another important difference is that, in (\ref{lulu}), $a+b+c$ is required to be negative,
which ensures that the volume in phase space contracts under the time evolution, as expected for a strange attractor.

Equations  equivalent to (\ref{lulu}) can be derived from 
a field theory Lagrangian if one reverses the sign in the kinetic term of the $z$ particle, {\it viz.} one is to
consider
a  model containing  three scalar fields $(X,Y,Z)$ with the following Lagrangian
\bea
L &=& -\frac12G_{ij} g^{\mu\nu}\partial_\mu X^i \partial_\nu X^j-V 
\nonumber\\
&=& \frac12 g^{\mu\nu} \left( \partial_\mu Z \partial_\nu Z   - \partial_\mu X \partial_\nu X -\partial_\mu Y \partial_\nu Y \right)
-V\ ,
\label{gigigi}
\eea
where $g_{\mu\nu}={\rm diag}(-1,1,...,1)$, $X^i=(X,Y,Z)$, $i=1,2,3$, with a target metric  $G_{ij}={\rm diag}(1,1,-1)$. 
The potential of the model is defined as follows:
\be
\label{gpo}
V= \frac12 (c Z-XY)^2-
\frac12 (aX+YZ)^2-\frac12 (bY+XZ)^2\ .
\ee

%With the focus on applications to isotropic and homogeneous cosmologies, we  look for 
Now consider solutions that have only time dependence. The Lagrangian becomes
\be
L = \frac12 \left( - \dot Z^2 + \dot X^2 +\dot Y^2    \right)
-V\ .
\ee
The potential (\ref{gpo}) can be derived from the superpotential
\be
\label{superw}
W= \frac12 \left(a X^2+b Y^2-c Z^2\right) + XYZ\ .
\ee
It satisfies the  relation (\ref{potiuno}),
\be
V= -\frac12 G^{ij}\partial_i W \partial_j W=\frac12 (c Z-XY)^2-
\frac12 (aX+YZ)^2-\frac12 (bY+XZ)^2\ ,
\ee
Using the general formulas (\ref{ecuuno}), we find  the following first-order equations,
\be
\label{primerorden}
\dot X = aX +YZ\ , \qquad 
\dot Y = b Y +XZ\ , \qquad
\dot Z = cZ -XY\ ,
\ee
which are identical to (\ref{lulu}). Here we have taken $\epsilon=1$; the opposite sign choice corresponds to the time-reversed solutions.

Thus there is a family of time-dependent solutions in the field theory that are governed by the system (\ref{lulu}). This system is chaotic in a large range of parameters.
The
basic dynamical properties of this chaotic system, including bifurcations, symmetries, periodic windows and Lyapunov exponents are investigated in detail in \cite{Lucheng}, so this
 analysis will not be reproduced here.
However, it is worth noting the important differences from the dynamical system  (\ref{didi}), which can be understood analytically. Consider a closed surface in phase space of volume ${\cal V}$.
%=\int_{v_0} dX\wedge dY\wedge dZ$. 
 The time derivative is given by
$$
\frac{\dot {\cal V}}{\cal V} =  \vec \nabla\cdot \dot {\vec X}=a+b+c\ ,\qquad \vec X=(X,Y,Z)\ .
$$
{} When $a+b+c<0$, the volume in phase space shrinks; the system is dissipative. Just like in the Lorenz attractor, the volume in phase space decays exponentially, showing the existence of an attracting set of zero volume.
In the system (\ref{didi}), one has $\dot {\cal V} = 0$ by virtue of
the identity $M_1+M_2+M_3=0$. This is already a sign that the system (\ref{didi}) is not chaotic.

 Choosing $a<0, b<0$ and $c>0$, the system (\ref{primerorden})
has five fixed points, one at the origin
$(X,Y,Z)=(0,0,0)$, and other four fixed points located at
\be
(X_0,Y_0,Z_0)\ ,\quad (X_0,-Y_0,-Z_0)\ ,\quad (-X_0,Y_0,-Z_0)\ ,\quad (-X_0,-Y_0,Z_0)\ .
\ee
where $X_0=\sqrt{-bc},\ Y_0=\sqrt{-ac},\ 
Z_0=\sqrt{ab}$.
If a fixed point is fully attractive in all directions, some trajectories will not be chaotic as they will fall into the attractive fixed point. Chaos is favored when all fixed points have repulsive directions. The fixed point at the origin
has a repulsive direction since one of the eigenvalues is positive, $c>0$.
Linearization around the other four fixed points give eigenvalues corresponding to the three roots of the cubic equation
\be
\lambda^3-(a+b+c)\lambda^2 +4 a b c=0\ .
\ee
With the choice of signs for $a,b,c$, with $a+b+c<0$, one eigenvalue is real and negative; the other two eigenvalues are complex conjugates. 
Taking $a,b<0$ and $c>0$, the complex eigenvalues have a positive real part, so they give rise to repulsive directions.
A strange attractor  then appears, as illustrated in  figures 1a,b.
%Figure 1 shows
%

%\begin{figure}[h!]
% \centering
% \includegraphics[width=0.45\textwidth]{Lu10_4.pdf}
% \caption{Trajectories for $a=-10$, $b=-4$, $c=20/7$. A double-scroll strange attractor.}
% \label{Luchaos}
% \end{figure}

%%%%%%%%%%%%%%%%%%%%%%%%%%%%%%%%%%%%%%%%%%%%%
\section{Cosmological model}
%%%%%%%%%%%%%%%%%%%%%%%%%%%%%%%%%%%%%%%%%%%%%

\subsection{Cosmological models with first-order equations}

Consider the following Lagrangian describing
Einstein gravity coupled to $n$ scalar fields $\phi^I$
with self-interactions,
\be
L=\frac12 \sqrt{-{\rm det}g} \left(2R- g^{\mu\nu} G_{IJ}(\phi) \partial_\mu \phi^I   \partial_\nu \phi^J-2V( \phi) \right)\ .
\ee
We shall study flat FLRW cosmologies of the form
\be
\label{flrwan}
ds^2= - e^{2\alpha\varphi } f^2 d\tau^2 +e^{2\beta \varphi} \left( dx_1^2+...+dx_{d-1}^2 \right)\ ,
\ee
with $\varphi=\varphi(\tau)$, $f=f(\tau)$, $\phi^I = \phi^I (\tau)$, and
\be
\alpha = (d-1)\beta 
\ ,\qquad \beta =\frac{1}{\sqrt{2(d-1)(d-2)}}\ .
\ee
The Einstein equations and the scalar field equations of the original Lagrangian then
reduce to the Euler-Lagrange equations of the effective Lagrangian
\be
\label{logro}
L_{\rm eff}=\frac{1}{2f} \left(- \dot \varphi^2+ G_{IJ}(\phi)  {\dot \phi^I}  { \dot \phi^J} \right)  -f e^{2\alpha\varphi} V( \phi)\ .
\ee
Even for very simple potentials, solving the  system of coupled second-order differential equations is very complicated.
Here we will be using a simple construction
where equations reduce to more tractable first-order differential equations.
The idea is to identify a class of  potentials that can be derived from a superpotential.
We  define the following superpotential:
\be
\WW = e^{\alpha \varphi} F( \phi )\ .
\ee
Then we choose the potential 
\be
\label{potro}
2e^{2\alpha\varphi}\, V\equiv (\partial_\varphi \WW)^2 -G^{IJ} \partial_I \WW \partial_J \WW =e^{2\alpha \varphi} \left(\alpha^2F^2-G^{IJ} 
\partial_I F\partial_J F\right)\ .
\ee
Note that the potential in general is not bounded from below.
This is a common feature in models arising from truncation of (gauged) supergravity/string theory (the simplest example being a negative cosmological constant). 

With the choice (\ref{potro}) for the scalar potential, a family of cosmological solutions
can be found by studying the following
first-order system:
\be
\dot\varphi =-f\partial_\varphi \WW=-\alpha f e^{\alpha\varphi} F,\qquad \dot{\phi _I }=f G^{IJ}\partial_J \WW= f e^{\alpha\varphi} G^{IJ}\partial_J F\ .
\ee
Choosing $f=e^{-\alpha\phi}$, the factor $e^{\alpha\varphi}$ cancels out and the first-order system takes the simple form
\be
\dot\varphi =-\alpha  F(\phi),\qquad \dot{\phi _I }=G^{IJ}\partial_J F(\phi)\ .
\ee
Note that an alternate choice of time,  corresponding to taking $f=e^{-\alpha\phi}/F$, identifies the time coordinate with $-\varphi/\alpha $.
However, for
our purposes it is more convenient
to choose $f=e^{-\alpha\phi}$. Then, the standard FLRW cosmological time $t$, defined by $dt=e^{\alpha\varphi}f d\tau$, coincides with $\tau$.

%%%%%%%%%%%%%%%%%%%%%%%%%%%
\subsection{The chaotic cosmological model}

Let us now apply this construction to  the three-scalar-field model studied in the previous section, coupled to gravity.
The lagrangian is
\be
\label{teorla}
L=\frac12 \sqrt{-{\rm det}g} \left(2R-G_{ij} g^{\mu\nu}\partial_\mu X^i \partial_\nu X^j-2V \right)\ ,
\ee
where $G_{ij}={\rm diag}(1,1,-1)$ as in (\ref{gigigi}) and the potential will be specified shortly.
The model contains a ghost field. 
Cosmological models with ghosts have been
extensively  used in cosmology to account for phantom phases where the
equation of state parameter $w$ is less than $-1$ (see {\it e.g.} \cite{Caldwell:2003vq,Nojiri:2015fia} and references therein). As usual, the presence of a ghost leads to well-known problems in the quantum theory, related to unitarity or vacuum stability (a discussion on phenomenological bounds  can be found in  
 \cite{Cline:2003gs}). In this work, of course, our goal is not the quantum consistency of the model, but rather  the understanding of the conditions under which classical deterministic chaos, such as that of generalized Lorenz models, can be incorporated into field theory.

We now consider the cosmological FLRW ansatz (\ref{flrwan}). The remaining Einstein's equations and scalar field equations can be derived
from the effective Lagrangian
\be
\label{laguna}
L_{\rm eff}=\frac{1}{2f} \left(- \dot \varphi^2-\dot Z^2 +\dot X^2+\dot Y^2\right)  -f e^{2\alpha\varphi} V\ .
\ee
The $f$ equation gives rise to the constraint:
\be
\label{contar}
\dot \varphi^2+\dot Z^2 -\dot X^2-\dot Y^2  = 2f^2 e^{2\alpha\varphi} V\ .
\ee
We shall choose the standard cosmological time, corresponding to the choice  $f=e^{-\alpha\phi}$.
We now consider a superpotential of the form
\be
\WW = e^{\alpha \varphi}\,  W\ ,
\ee
where $W$ given by (\ref{superw}). The potential is then defined by
\bea
 2V &=& \alpha^2 W^2 + (\partial_Z W)^2- (\partial_{X} W)^2-(\partial_{Y} W)^2
\nonumber\\
&=& \frac{\alpha^2}{4}\left( a X^2+b Y^2-c Z^2 +2 XYZ\right)^2
\nonumber\\
&+& (c Z-XY)^2-
 (a X+ YZ)^2- (bY +ZX)^2\ .
\label{elpoti}
\eea
The parameters $a,b,c$ represent self-interaction couplings for the three scalar fields.
%The resulting potential has unstable directions at infinity.
Remarkably, with the choice $f=e^{-\alpha\varphi}$, we  find exactly
the first-order system (\ref{primerorden}), supplemented with the additional equation
\be \label{ffii}
\dot \varphi = -\alpha W\ .
\ee
The solutions to these equations  solve the second-order equations of the original theory (\ref{teorla}) and the constraint  (\ref{contar}).
One can first solve  (\ref{primerorden}) for $X,Y,Z$ and then 
substitute the solution into $W$, see (\ref{superw}).
The cosmological metric is then obtained by integrating (\ref{ffii}).

Thus  trajectories in the three-dimensional space $(X,Y,Z)$ 
are governed by the same autonomous system (\ref{primerorden}) as in the generalized Lorenz system studied by \cite{Lucheng}.
Chaotic trajectories appear
for a wide range of couplings. 
For example, the choice 
$$
a=-10\ , \qquad b=-4\ ,  \qquad  c=-\frac{ab}{a+b}\ ,
$$
studied in  \cite{Lucheng}, gives rise to
  a double-scroll strange attractor with a spectrum of Lyapunov exponents \cite{Lucheng}
$\lambda_1\approx 1,17$, $\lambda_2=0$ and $\lambda_3\approx -12.3$ and Lyapunov dimension $d_L\approx 2.1$ for initial value $(1,1,1)$ (other choices of $c$ can generate  a 4-scroll chaotic attractor).
Note that the motion is bounded, despite  the potential having unstable directions at infinity.

%(here Let us consider parameters $a=-10$, $b=-4$, $c=-ab/(a+b)=20/7$).

\begin{figure}[h!]
 \centering
 \begin{tabular}{cc}
\includegraphics[width=0.4\textwidth]{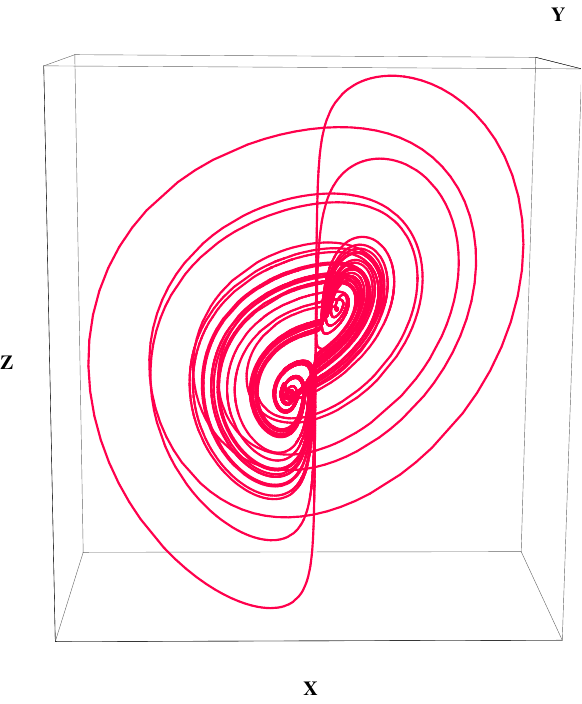}
 &
 \qquad \includegraphics[width=0.4\textwidth]{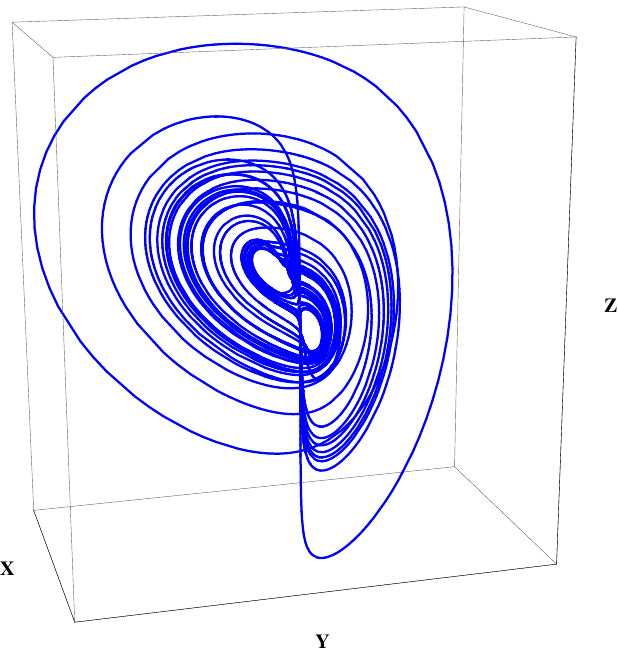}
 \\ (a)&(b)
 \end{tabular}
 \caption
 {The double-scroll strange attractor for two different sets of parameters (a)   $a=-10$, $b=-4$, $c=20/7$. 
 %$a=-2$, $b=-5$, $c=10/7$.
  (b) $a=-2$, $b=-4$, $c=4/3$.
 }
 \label{Luparam}
 \end{figure}

There is a range of parameters $a,b,c$ where there are no chaotic trajectories. 
For example, when $a+b+c>0$,  trajectories typically go to infinity.
On the other hand, when $a,b,c$ are all negative, the fixed point at the origin is attractive  and solutions can approach at a scaling solution at this attractive fixed  point. 
These alternative regimes give rise to more
familiar cosmologies. 
However, our main focus here is to  investigate the properties of the cosmological solutions associated with the chaotic trajectories, since this appears to be a new behavior, which, to our knowledge, has not been considered before.

\begin{figure}[h!]
 \centering
 \begin{tabular}{cc}
\includegraphics[width=0.4\textwidth]{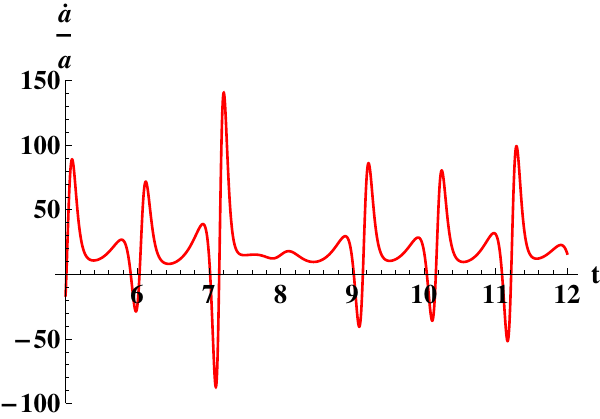}
 &
 \qquad \includegraphics[width=0.4\textwidth]{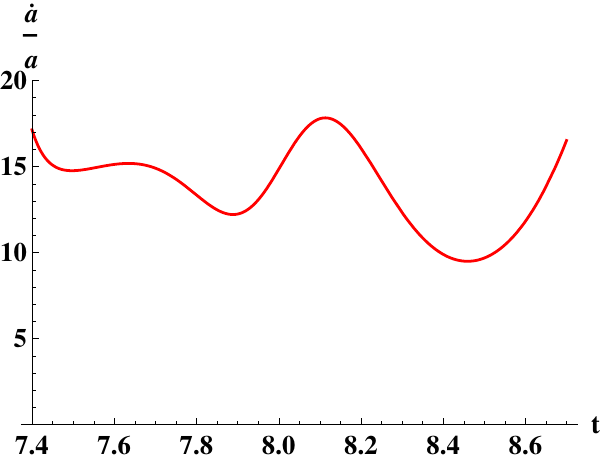}
 \\ (a)&(b)
 \end{tabular}
 \caption
 {(a) The expansion velocity given by the Hubble parameter $H(t)=\dot a(t)/a(t)$ for a solution in the $d=4$ theory with couplings $a=-10$, $b=-4$, $c=20/7$. The cosmological evolution shows periods of rapid expansion and contraction, alternating with steady periods where $w\approx -1$.
  (b) Same plot enlarged, showing in detail a period  of accelerated expansion where $w\approx -1$
  ({\it cf.} fig. \ref{wpara}).
 }
 \label{hubblep}
 \end{figure}

The scale factor is $a= e^{\beta \varphi}$.
The universe expands if $\dot a = \beta e^{\beta \varphi}\dot \varphi >0$. Since $\dot \varphi =-\alpha W$, periods of expansion occurs when the trajectories go through a region where $W<0$.
Along the chaotic trajectories, $W$ fluctuates taking positive and negative values. This is shown in fig. \ref{hubblep}a.
Thus the universe undergoes periods of contraction and expansion.
Surprisingly, the cosmological evolution also shows  periods with steady behavior. Let us now discuss this feature in more detail.

The expansion is accelerated when $\ddot a>0$. From the relation
\be
R_{00}=-(d-1)\frac{\ddot a}{a}\ ,
\ee
one sees that universe undergoes an accelerated expansion in the regions where $W<0$ and
$R_{00}<0$.
{}From the Einstein's equations, we find
\be
R_{00}=\frac12 \left( \dot X^2+\dot Y^2-\dot Z^2 -\frac{2}{d-2}V\right) \ .
\ee
Upon using the constraint, this reduces to
\be
R_{00}=\frac12 \left( \dot\varphi^2  - \frac{2(d-1)}{d-2}V\right)= \frac{d-1}{4(d-2)}  \left( W^2  -4V\right)
\ .
\ee
where $\dot \varphi=-\alpha W$ has been used.

%\be
%R_{00}= \frac{d-1}{4(d-2)}  \left( -\frac{W^2}{d-2}  +2G^{ij}\partial_i W\partial_j W\right)
%\ee
\bigskip

\begin{figure}[h!]
 \centering
 \begin{tabular}{cc}
\includegraphics[width=0.4\textwidth]{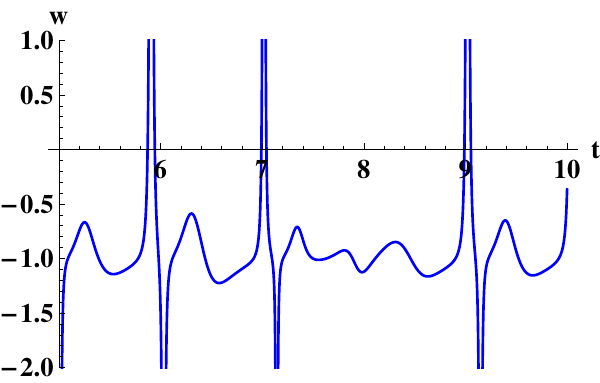}
 &
 \qquad \includegraphics[width=0.4\textwidth]{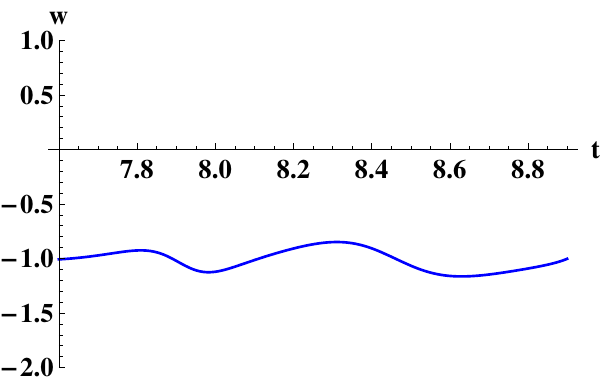}
 \\ (a)&(b)
 \end{tabular}
 \caption
 {(a) $w$ as a function of time for the same trajectory
 as in fig. \ref{hubblep}. The expansion is accelerated in the region $w<-1/3$, and it approaches a de Sitter-like expansion when $w\approx -1$. There are singular points where $W$ vanishes, corresponding to $\dot a=0$. (b) 
 Same plot enlarged in a region where $w\approx -1$.
  }
 \label{wpara}
 \end{figure}

Following \cite{Russo:2022pgo}, we can study the evolution of the equation of state along a given trajectory.
The equation of state is given by $p=w\rho$, where $p$ is the matter pressure and $\rho $ is the energy density.
Their explicit expression is found by  
computing
$T_{00}=-g_{00}\rho\ ,\ \ T_{ij} = g_{ij} p\ $.
For the present model,
\be
\rho=\frac{1}{4}\left(  \dot X^2+\dot Y^2-\dot Z^2+2V\right)\ ,\qquad
p=\frac{1}{4}\left(  \dot X^2+\dot Y^2-\dot Z^2-
2V\right)\ .
\ee
Using the constraint
\be
\dot X^2+\dot Y^2-\dot Z^2 =
\dot\varphi^2  - 2V\ ,
\ee
the density and pressure reduce to
\be
\rho=\frac{1}{4} \dot\varphi^2\ ,\qquad
p=\frac{1}{4}\left(  \dot\varphi^2 - 4V\right)\ .
\ee
Hence
\be
w=1-\frac{4V}{\alpha^2 W^2}\ .
\ee
We note that a de Sitter-like phase with $w \approx -1$ arises in the region where
$2V\approx \alpha^2 W^2$. From (\ref{elpoti}), we see that this occurs 
%precisely 
near the extrema of the superpotential, where $|\partial_i W|\ll |W| $, that is, when the trajectory passes through the neighborhood of a fixed point.
In this case the kinetic energy is small and FLRW spacetime approximates de Sitter space during some time (see figs \ref{wpara}a,b).

%\section{Discussion}
\medskip

In conclusion,
here we studied a cosmological model with three self-interacting scalar fields, one of them with negative kinetic term.
The model exhibits a family of solutions represented by a strange attractor, which is closely related to the classic Lorenz attractor, whose fascinating dynamical behavior has been extensively studied in chaos theory.
The three-dimensional autonomous system is almost identical to the system (\ref{didi}), describing (non-chaotic) classical solutions of ${\cal N}=1^*$ $SU(N)$ super Yang-Mills theory, differing from it  in the sign of a quadratic term. 
In the gauge theory, the sign flip would be achieved if the $SU(N)$ gauge group is changed by a non-compact group containing $SU(1,1)$,
again leading to a ghost scalar field, which appears to be the root of 
 the emergence of chaotic behavior.

\smallskip

The `strange-attractor' universe might model the physics near the cosmological singularity. We have seen that the cosmological evolution is subject to strong fluctuations over a period of time, but, soon after, the universe reaches a steady behavior with a  de Sitter-like expansion. 
It would be interesting to see if
an ansatz based on a more general  metric could  lead
to similar chaotic behavior, without the need to add a ghost field.
More generally, it would be important to understand the general conditions under which
strange attractors can appear in multiscalar cosmological models.

\subsection*{Acknowledgments}
We would like to thank Paul Townsend for useful discussions and comments.
We acknowledge financial support from a MINECO
grant PID2019-105614GB-C21.

\end{document}